Superconducting Accelerators for High-Power X-ray production

Thomas K. Kroc, PhD

Fermi National Accelerator Laboratory

September 29, 2021

**Forward**

To date, linear accelerators (linacs) as electron sources used to produce ionizing radiation for industrial purposes have been limited to less than 100 kW. When the electron beam is used directly, this is sufficient for most potential applications. However, when the electron beam is used for the production of photons (x-rays) which are then to be used in an application, this is not sufficient to compete with other sources of photons (gamma rays from cobalt-60). This paper will discuss a compact superconducting RF (CSRF) accelerator system that relies on emerging technologies that will be able to produce electron beam powers into the 100s of kW with efficiencies much better than present linacs. The focus is to produce x-ray beams for medical device sterilization to provide alternatives for the present use of cobalt-60 for this purpose.

**Table Of Contents**

## 1. Introduction

Approximately 50% of single use medical devices are sterilized with gamma rays from the decay of Co-60. Worldwide, there is over 400 MCi of cobalt installed in facilities in over 40 countries. However there is permitted capacity of over 600 MCi for those facilities [1]. The medical device industry is growing at 7% per annum. This number does not include the recent impact of COVID-19. The sterilization market is very tight and sterilization capacity shortages are looming, The production of cobalt-60 is presently behind market demand by 5% [1] and given the cycle times of cobalt production, it may take a few years for production to catch up. Additionally, there are radioisotope security concerns that lead some to press for a reduction in our dependence on radioisotopes.

As can be seen in Fig. 1, the penetration characteristics of X-rays produced by a 7.5 MeV electron beam is very similar to the penetration of the gamma rays from cobalt-60. This means that x-rays can be a direct replacement for gamma sterilization of medical devices.

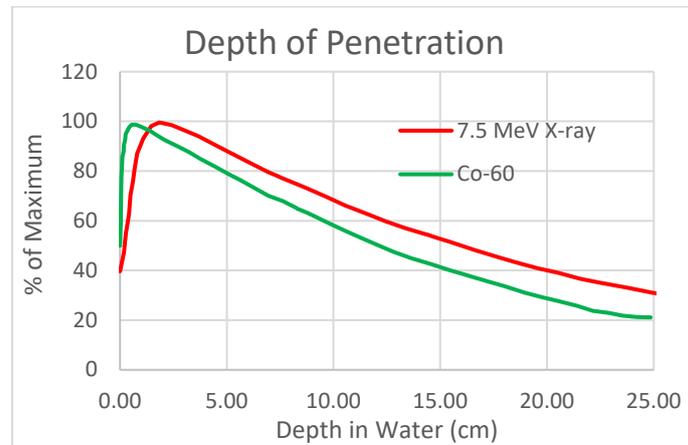

*Figure 1 Comparison of the depth of penetration of photons from cobalt-60 decay and Bremsstrahlung x-rays from a 7.5 MeV electron beam.*

Superconducting accelerators offer high-power and efficient sources of ionizing radiation to effectively sterilize all products that are presently sterilized using gamma rays from cobalt-60. This potential comes from two areas: 1) the efficiency of the accelerating system itself, and 2) the efficient utilization of energy to produce the RF energy needed to drive the accelerator.

One megacurie (MCi) of cobalt-60 produces about 15 W of power in the gamma rays that are released. Due to the inefficiency of the Bremsstrahlung process, it requires about 120 kW of electron beam onto a high-Z target to produce the same amount of power in the resulting x-rays.

Normal conducting accelerators (which operate at room temperature) produce inductive currents in the conducting, inner surface of the accelerating cavities. These currents result in the generation of heat and dissipation of energy due to the resistance of the cavity material (typically copper). The amount of power lost can be greater than the power of the beam of the accelerated particles. In a superconducting system, the resistance of the cavity surface is negligible resulting in all the supplied power being utilized in the power of the resulting beam.

## 2. System Efficiency [2]

An economic analysis of the cost of gamma and x-ray facilities suggests that long term costs for an x-ray facility are less than for gamma when the effective capacity is greater than 1–2 MCi. [3]. Despite this, any economic advantage is not significant enough to overcome other expenses of revalidating existing devices. This results in ambivalence in modality from a cost perspective for new devices and provides little incentive to revalidate existing devices in a new modality. Improving the electrical efficiency of accelerator systems would improve this and there is considerable room for improvement in this area.

Here, we examine the factors that determine the efficiency of an accelerator system. The overall efficiency of industrial superconducting accelerators, $\eta_{acc}$, is determined by the following factors [4]:

- Efficiency $\eta_{AC \to DC}$ of the conversion from AC to DC; typically, the efficiency of transformers and rectifiers is higher than 95%;

- Efficiency of the RF source $\eta_{DC \to RF}$ including additional power for the gun heater and electromagnets for vacuum tubes (this power is typically small compared to other parameters);
- The need for equipment cooling. A chiller adds additional power overhead $\xi_{chill}$ that is approximately 15% of the power dissipated in the RF source;
- Input power determined by the RF source gain K (in dB); the input power is ~0.001–5% of the output RF power, depending on type.
- Transmission line efficiency $\eta_{trans}$; for industrial linacs, where the RF source is typically placed near the accelerator, it is close to 100%;
- Accelerator cavity efficiency $\eta_{RF \to beam}$; for a superconducting RF (SRF) linac operated in the continuous mode (CW) regime it may be close to 100% if the cavity is matched to the line. Note that in a pulsed regime, $\eta_{RF \to beam}$ is smaller because of the filling time, RF source timing off-set, the modulator off-set, etc.;
- Efficiency of the electron injector $\eta_{inj}$,
- Efficiency of the cavity cryogenic cooling $\eta_{cooler}$, which in turn is determined by the RF losses $P_{loss}$ in the cavity and cooling efficiency. For cryo-plant (2K) it is about 1–1.3x$10^{-3}$, for cryo-cooler (~ 4 K) it is 1–1.5x$10^{-4}$.

The energy consumption diagram for CW accelerator is shown in Fig. 2.

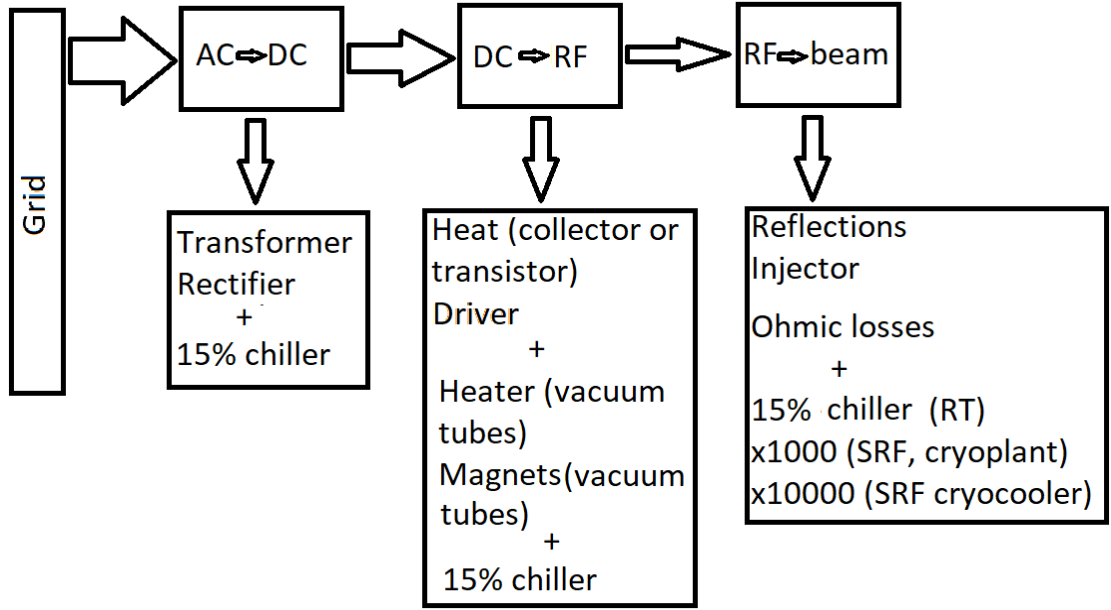

*Figure 2 Power consumption diagram for CW accelerator.*

The total accelerator system efficiency may be calculated using the following formula:

$$\eta_{acc}^{-1} = 1 + (1/\eta_{RF \to beam} - 1)(1 + \xi_{chill}) +$$
$$+ \frac{\left[\left(\frac{1}{\eta_{DC \to RF}} - 1\right)(1 + \xi_{chill}) + (1 - \eta_{trans}) + \frac{1}{\eta_{DC \to RF}}\left(\frac{1}{\eta_{AC \to DC}} - 1\right) + 10^{-0.1K}\right]}{(\eta_{acc}\eta_{trans})}$$
$$+ \frac{P_{inj}\left(\frac{1}{\eta_{inj}} - 1\right)}{P_{beam}} + \frac{P_{loss}}{P_{beam}\eta_{cooler}}$$

The overall efficiency of an industrial accelerator is determined to a high degree by the efficiency of the RF sources, especially for SRF accelerators. If we consider 650 MHz, 10 MeV, 250 KW linac based on the $Nb_3Sn$ technology, we may expect the unloaded quality factor $Q_0$ of up to $3 \times 10^{10}$ at 5–6 MeV/m at 4.5 K, and R/Q of up to 670 Ohm/m. It means that for a cavity length of ~1.6 m, the total surface Ohmic loss is 3 W. If the cryo-cooler efficiency is $\gtrsim 10^{-4}$, the total power necessary for the cooling system is $\lesssim 40$ kW. If one assumes that the power necessary to excite the RF source is small and that the losses in the electron injector are small, the formula is simplified:

$$\eta_{acc}^{-1} = 1 + \left(\frac{1}{\eta_{DC \to RF}} - 1\right)(1 + \xi_{chill}) + \frac{1}{\eta_{DC \to RF}}\left(\frac{1}{\eta_{AC \to DC}} - 1\right) + \frac{P_{loss}}{P_{beam}\eta_{cooler}}$$

In this case, the overall efficiency of the complete CSRF system is determined by the efficiency of the RF source as shown in Fig. 3.

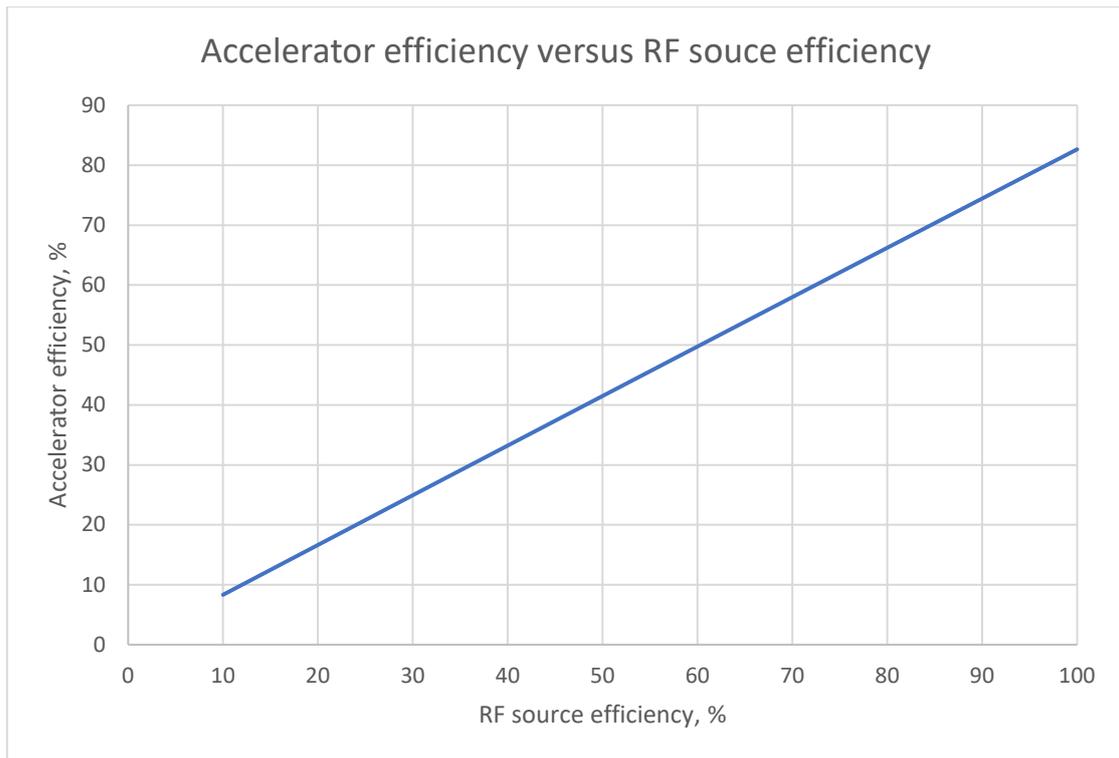

FIG. 3. Total efficiency $\eta_{acc}$ of the 10 MeV, 250 kW CWSRF accelerator versus RF source efficiency $\eta_{DC \rightarrow RF}$.

Given the state of climate change, it behooves us to use energy as efficiently as possible.

## 3. Enabling Technologies for Industrial Superconducting RF Accelerators

The efficiency of superconducting accelerators comes from the integration of 4 enabling technologies. Three of these minimize the introduction or generation of heat into the system and one removes that heat from the system without the need for liquid cryogens.

Minimize heat generation

- Integrated electron source
- Low heat-loss RF coupler
- $Nb_3Sn$ coated cavities

Efficiently remove heat

- Conduction cooling & commercial cryocoolers

The integrated electron source does not use a Low Energy Beam Transport (LEBT) line. Rather the thermionic cathode is positioned in a small opening in the superconducting cavity. Despite the high temperature of the cathode (~ 1050 C), its small size and careful design limit the amount of heat introduced into the system to 0.3 W.

The low heat-loss coupler is also a connection from the cryogenic nature of the superconducting cavity to the ambient environment. Thin walls and an intermediate temperature intercept limit to conducted heat to 1 W. Additionally, the use of solid copper for the electromagnet structures eliminates the risk of flaking from electroplated elements. These flakes can spoil the conductance of the coupler.

The most critical aspect of the superconducting design is the use of a $Nb_3Sn$ coating on the inner surface of the accelerating cavity. Superconducting cavities have frequency dependent currents that are induced in the walls of the cavity. Our choice of 650 MHz reduces the induced currents to a manageable level. This coating doubles the critical temperature of the surface which allows operation at ~4K instead of 2K. Operating at 4K allows the use of cryocoolers to remove the heat from these currents. A critical measure of the performance of superconducting cavities is the quality factor $Q_0$ which is the ratio of the initial stored energy to the energy lost in each oscillation of the RF. We have demonstrated cavities with a $Q_0$ of greater than $1 \times 10^{10}$ with an average electric field gradient of about 7 MV/m. This is a very conservative gradient compared to accelerators used in discovery science and is a key element in the design to provide robustness for use in industrial environments.

Commercial cryocoolers are now able to dissipate 2–2.5 W of heat at 4K, which is enabled by the $Nb_3Sn$ coating as noted above. These coolers operate without liquid cryogens and the large infrastructure needed to produce and maintain the cryogens. To efficiently move the heat from their origin, a method has been developed [5–9] to conductively transfer the heat to the cryocoolers. This uses high purity aluminum and required a thorough investigation of proper design, sizing, and fastening to attain a fraction of a degree temperature change across the transfer path.

## 4. 20 kW Prototype

To validate the compact SRF concept, we are in the process of assembling a 1.6 MeV, 20 kW prototype that incorporates all of the enabling technologies noted above with funding from the National Nuclear Security Administration. As shown in Fig. 4, the cryostat is about 1 meter long and 1 meter in diameter. Two cryocoolers (one shown) will provide the necessary cooling capacity. It has one full-length accelerating cavity and one partial cavity. The partial cavity matches the beam emitted from the electron source to the full cavity. We refer to this combination as a 1.5 cell cavity. This is the minimal system to illustrate the successful integration. Assembly should be completed in late 2023. It is possible that this device itself may have commercial applications.

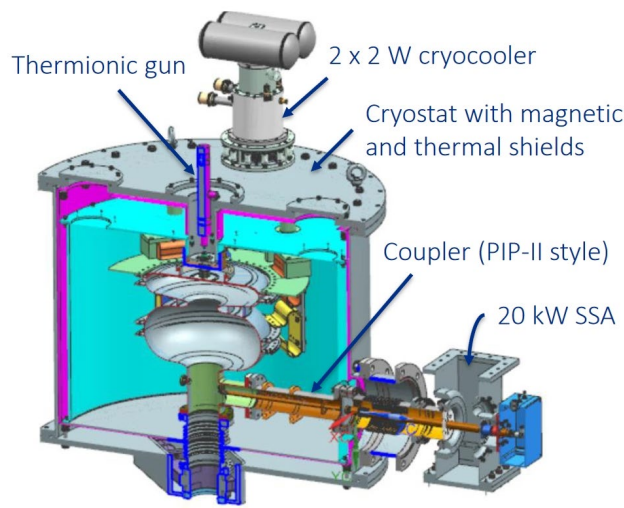

FIG. 4  Diagram of 1.6 MeV, 20 kW prototype with 1.5 cell $Nb_3Sn$ accelerating cavity, integrated electron gun, and low heat-loss RF coupler.

## 5. 200 kW + systems

With the successful demonstration of the 20 kW prototype, the design will be ready for the production of a commercial scale system that can produce the equivalent of 2 MCi or more. In a previous effort we have developed a digital design of such a system. A preliminary illustration of the system is shown in Fig.5. While an actual system will not be as compact as shown here it will still be very compact. This system utilizes 5.5 cells to achieve an electron beam energy of 7.5 MeV while still using the conservative gradient noted previously. The 5.5 cells generate 6.4 watts of heat compared to the 1.5 watts in the 1.5 cell cavity. This system will require 4 cryocoolers, so the gold structure will require two racks. Two 100 kW magnetrons would be housed in the light blue rack. Controls and vacuum supplies would occupy the dark rack. This would be all that is needed to supply a scanning horn with a Bremsstrahlung converter as is presently used in present x-ray sterilization applications.

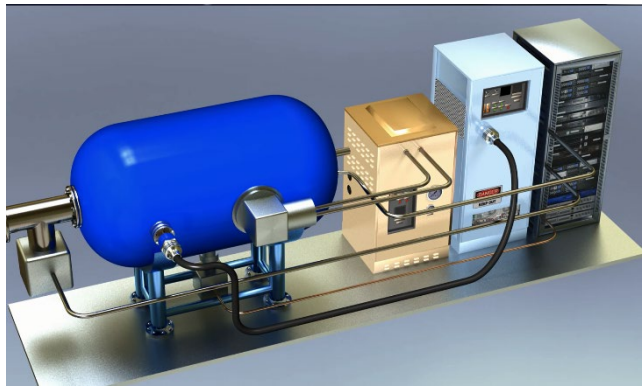

*FIG. 5. Conceptual illustration of 200 kW compact SRF accelerator system including cryocooler compressor, control racks, and magnetron RF supply.*

## 6. RF Needs [2]

In order to fully realize the efficiency potential of the superconducting system, efficient RF power supplies are required. Solid state power supplies are becoming a favorite of the market. However, the electrical efficiency of these supplies is less than 50%. Magnetrons are quite simple, inexpensive and have electrical efficiencies of 80% or more. However, the performance requirements for currently available commercial magnetrons may not be suitable for these high-performance applications, particularly the expected lifetime, which can be quite variable at present. To that end, we have identified a development plan to improve the performance of magnetrons while still being able reap the economy of these devices.

A major issue that may inhibit the use of magnetrons for industrial accelerators is their low life span. Magnetron tubes are inexpensive, so the cost of replacement is not a concern in present commercial applications such as the process heating industry (drying wallboard and lumber). For an industry such as medical device sterilization, the concern would be process interruption on a frequent and irregular basis. Longer lifetime and a more dependable mean-time-to-failure may be necessary for adoption of this power source. The main reasons for the present state of tube lifetime are:

- Anode sputtering of the cathode material;
- Cathode bombardment by backward electrons.

General measures which may be taken to address these issues are:

- Active vacuum pumping of the magnetron;

- Electron dynamics optimization.

Attention to electron dynamics may also improve the magnetron efficiency. Recent investigations show [10, 11] that magnetron efficiency may be improved together with lifetime extension by operating in a sub-critical regime.

While we have concluded that our next efforts should be focused on completing the integration of a complete CSRF system, we have outlined a program to address these issues of magnetron lifetime and efficiency.

**The program:**
1. Use a modern 3D simulation code to understand in detail the beam dynamics of a magnetron.
2. Benchmark the code. This improved and benchmarked code will strengthen the RF industry allowing better designs of the magnetron for different applications – scientific, industrial, civil, and military.
3. Finally, it would be possible to optimize the magnetron design to improve its longevity and efficiency and optimize various operation regimes. Different options could be explored, like 2D harmonic cavities, different types of cathodes including the newly developed Nanocomposite Scandate Tungsten cathodes [12].

The goal would be to achieve an efficiency of more than 85% with tube lifetime of ~50,000-80,000 hours.

## 7. Conclusion

The need for new sources of ionizing radiation for medical device sterilization is acute. New technology is arriving to provide higher-power, more economical x-ray sources. These x-ray sources can provide direct alternatives to gamma ray sources. The compact superconducting accelerator system described here utilizes four enabling technologies to provide a new generation of linac sources.